\documentclass[12pt,preprint]{aastex}
\usepackage{amsmath,amssymb,rotating,natbib} %%%%
\newcommand{\mapiii}{MAPPINGS  \textsc{iii}}

\newcommand{\mum}{\ensuremath{\mu\mbox{m}}}
\newcommand{\pccm}{\ensuremath{\,\mbox{cm}^{-3}}}

\catcode`\@=11
\def\gapprox{\mathrel{\mathpalette\@versim>}}
\def\lapprox{\mathrel{\mathpalette\@versim<}}
\def\@versim#1#2{\lower2.45pt\vbox{\baselineskip0pt\lineskip0.9pt
     \ialign{$\m@th#1\hfil##\hfil$\crcr#2\crcr\sim\crcr}}}
\catcode`\@=12
\newcommand{\hii}{H\,{\sc ii}}

\newcommand{\heii}{He\,{\sc ii}}

\newcommand{\cii}{C\,{\sc ii}}
\newcommand{\ciii}{C\,{\sc iii}}
\newcommand{\civ}{C\,{\sc iv}}

\renewcommand{\ni}{N\,{\sc i}}
\newcommand{\nii}{N\,{\sc ii}}
\newcommand{\niii}{N\,{\sc iii}}

\newcommand{\nv}{N\,{\sc v}}
\newcommand{\oi}{O\,{\sc i}}
\newcommand{\oii}{O\,{\sc ii}}
\newcommand{\oiii}{O\,{\sc iii}}

\newcommand{\ovi}{O\,{\sc vi}}

\newcommand{\neii}{Ne\,{\sc ii}}
\newcommand{\neiii}{Ne\,{\sc iii}}

\newcommand{\nev}{Ne\,{\sc v}}
\newcommand{\nevi}{Ne\,{\sc vi}}

\newcommand{\sii}{S\,{\sc ii}}

\newcommand{\arii}{Ar\,{\sc ii}}
\newcommand{\ariii}{Ar\,{\sc iii}}

\newcommand{\arvi}{Ar\,{\sc vi}}
\begin{document}
\title{Dusty, Radiation Pressure Dominated Photoionization. \\
II. Multi-Wavelength Emission Line Diagnostics for Narrow Line Regions}

\author{Brent A. Groves, Michael A. Dopita, \& Ralph S. Sutherland}

\affil{Research School of Astronomy \& Astrophysics,
Institute of Advanced Studies, The Australian National University,
Cotter Road, Weston Creek, ACT 2611
Australia}
\email{bgroves@mso.anu.edu.au, Michael.Dopita@anu.edu.au,
ralph@mso.anu.edu.au}

\begin{abstract}
Seyfert narrow line region (NLR) emission line ratios are remarkably
uniform, displaying only $\sim$0.5 dex variation between galaxies, and
even less within an individual object. Previous photoionization and
shock models of this region were unable to explain this observation
without the introduction of arbitrary assumptions or additional
parameters. Dusty, radiation pressure dominated photoionization models
provide a simple physical mechanism which can reproduce this spectral
uniformity between different objects. In the first paper of this
series we described this model and its implementation in detail, as
well as presenting grids of model emission lines and examining the
model structures. Here we explore these models further, demonstrating
their ability to reproduce the observed Seyfert line ratios on
standard line diagnostic diagrams in both the optical and UV. We also
investigate the effects that the variation of metallicity, density and
ionizing spectrum have upon both the new paradigm and the standard
photoionization models used hitherto. Along with the standard
diagnostic diagrams we provide several new diagnostic diagrams in the
UV, Optical and IR. These new diagrams can provide further tests of
the dusty, radiation pressure photoionization paradigm as well as
being used as diagnostics of the metallicity, density and ionizing
spectrum of the emission line clouds.
\end{abstract}


%%%%%%%%%%%%%%%%%%

\section{Introduction}

The emission lines of active galaxies have often been used in
conjunction with models to constrain the physical and ionization
structure of the emitting regions.  In particular ratio diagrams or
line diagnostic diagrams prove to be an excellent visual aid in
interpreting the emission line data.  First used systematically by
\citet{BPT81}, line diagnostic diagrams take the observed ratios from
emission line galaxies and create a two-dimensional classification
scheme which is better able to differentiate between excitation
mechanisms and determine other properties of emitting regions such as
density or chemical abundance in the gas phase.  The line diagnostic
diagrams by \citet{VO87} are particularly useful as they involve
ratios of lines which are not greatly separated in wavelength and so
minimize the effects of differential reddening by dust within the
emission line region. These diagrams are capable of distinguishing
three different groups of emission line galaxies: those excited by
starbursts and two classes of object excited by an active nucleus -
the Seyfert narrow line regions (NLRs) and the low ionization nuclear
emission-line regions (LINERs).  These diagrams are additionally
interesting in that they show that the emission from NLRs is
remarkably uniform, with only $\sim$0.5 dex variation between Seyferts
and less within individual galaxies. This uniformity of the spectral
properties has since been confirmed in much larger samples
\citep[eg][]{VeronCetty2000}.

The use of line diagnostic diagrams (LDDs) is not limited to
distinguishing the excitation source of active galaxies. They can also
help us understand the details of the physical processes going on
within the ionized nebulae by comparing observations with the
predictions of both photoionization and shock models
\citep{Evans85,do:su95}.  The best fitting models can, in turn,
indicate the physical parameters and excitation mechanisms of the NLR.

Such comparisons have led to the acceptance of both of the mechanisms
for excitation in narrow line regions.  However, excitation by fast,
radiative shocks \citep{do:su95, do:su96} appears mostly in the
extended NLR (ENLR) associated with radio galaxies and in LINERs, and
is applicable only for a few Seyfert galaxies.  The photoionization
models, and variations thereof
\citep[e.~g.][]{KomSch97,bws96,bwrs97,Baldwin95, Ferg97}, are able to
reproduce the Seyfert observations with only a few failings. The main
problem with the standard photoionization models is that they are
unable to provide the observed uniformity in emission line ratios
without making arbitrary (and possibly unphysical) assumptions.  The
less than
0.5 dex variation in observed line ratios requires an approximately
constant ionization parameter of $U\sim10^{-2}$, where the ionization
parameter is a measure of the number of ionizing photons against the
hydrogen density ($U = S_{\star}/\mathrm{n}_{\mathrm{H}}c$).  Since
there is no physical reason why the ratio of photon density to gas
density should be constant within a given Seyfert, or from one Seyfert
to another, such a result is puzzling to say the least. Furthermore,
many Seyferts show strong coronal lines of highly ionized species
which can only be produced in a plasma with $U \gapprox 1.0$. So why
do low and intermediate excitation species indicate $U \sim 10^{-2}$?

In order to account for these failings of the standard models we have
proposed a new paradigm for the photoionization of the NLR clouds,
that of dusty, radiation pressure dominated photoionization
\citep[hereinafter DG02]{DG02}.

We have demonstrated previously that this model provides a
self-consistent explanation for the remarkable similarity between the
emission line spectra of NLR. The radiation pressure acting upon the
dust provides a simple controlling factor for the moderation of the
density, excitation and surface brightness of the emission line
region. This limits the models of the low- and intermediate-excitation
line ratios at high ionization parameters to $\sim$0.5 dex in the line
diagnostic diagrams, but allows the coexistence of high ionization
parameter and low density regions within a single model.

Following the introduction of this model, we gave details on how this
model was implemented within the photoionization \& shock code
\mapiii\ and a grid of UV, Optical and IR emission lines, covering a
range in metallicity, density, power-law ionizing continuum and
photoionization parameter \citep[Paper 1 of this series]{GDS03}. We
also examined the physical structure of the dusty model and how this
varied with the different input parameters.

We continue this work here, presenting line diagnostic diagrams for
both our dusty model and the standard photoionization model derived
from the line ratios given in Paper 1.
 
%%%%%%%%%%%%%%%%%%%%%%%%%%%%%%%

\section{Dusty, Radiation Pressure Dominated Models\label{sec:radpress}}

The inclusion of dust into photoionization models affects the final
emission spectrum in several ways. As well as simply absorbing EUV
radiation and competing with Hydrogen for the ionizing photons, dust
affects the temperature structure of the NLR clouds through the
process of photoelectric heating.  

In order to be physical, an isobaric photoionization model must
include the effects of radiation pressure. The force of radiation can
be imparted to both the gaseous medium and dust, and
results in a radiation pressure gradient. Since the grains are charged
and are therefore locked to the plasma by coulomb forces,
the radiation pressure gradient on dust
results in a gas pressure gradient identical in size. Standard
photoionization models are isochoric and therefore cannot take this
effect into account.  To demonstrate these effects we have run
the standard isochoric model and the new dusty, radiation pressure
dominated model over a large set of input parameters.

With simple calculations it is easy to show that at an ionization
parameter of $\log U \sim -2$, dust begins to dominate the opacity of
the ionized cloud and hence the radiation pressure. It is also around
this value of the ionization parameter that radiation pressure
developed at the ionization front of the NLR cloud becomes comparable
with the gas pressure.  Therefore at high ionization parameters, the
pressure in the ionized gas, and hence density (since the electron
temperature is always $\sim 10^4$K), is determined by the external
ionization parameter, $U_0$ and the local ionization parameter becomes
independent of the external ionizing flux. The result of this is that
at high ionization parameter the emission line spectrum of the low-
and intermediate ionization species is effectively independent of the
external ionization parameter. This independence was illustrated
through the example line diagnostic diagrams (LDDs) given in DG02 and
is demonstrated again in this paper.

Once the exciting mechanism of the NLR of the active galaxies has been
clearly identified, the diagnostic line ratio plots will
also be useful in defining the physical conditions within the
NLR. To facilitate this, we have run a series of the dusty,
radiation pressure dominated models covering a range in density,
metallicity, power-law index of the ionizing spectrum, and
photoionization parameter. To provide a comparison we also present a
series of dust-free, standard photoionization models covering the same
range of parameters. We described the implementation of these models
in depth in Paper 1 of this series \citep{GDS03}, where we gave the
emission line strengths from these models in a sequence of tables
covering the parameter space. These tables were used to construct
the diagnostic plots presented here.

\subsection{Parameter Space}\label{sec:param}

The parameters were chosen to cover the range of
values that would be reasonably expected to be found in the NLR of
Seyfert galaxies. The spread in parameter space allowed an examination
of how each parameter affected the models, and was great enough that
the resulting line emission from each model was distinct.

Three Hydrogen number densities ($\mathrm{n}_\mathrm{H}$) were
modeled; n$_\mathrm{H} = 10^2,~10^3$ and $10^4$ \pccm. For the
standard isochoric (constant density) models such modeling is
straightforward. However for the isobaric, dusty, radiation pressure
dominated models the concept of a single density for the nebula no
longer holds. Thus we have used the region near the ionization front,
where n$_{\mathrm{HII}}/$n$_{\mathrm{HI}} \sim 1.0$, as the fiducial
point at which to set the density.  It is in this region that the
density sensitive lines like [\sii]$\lambda \lambda 6717,30$\AA\ or
[\oii]$\lambda\lambda 3727,9$\AA\ are at their strongest and so this
region defines the effective electron density that line observations will
measure.

Five Metallicities were examined in the models; $0.25
\mathrm{Z}_\odot,~ 0.5 \mathrm{Z}_\odot,~ 1 \mathrm{Z}_\odot,~ 2
\mathrm{Z}_\odot$ and $4 \mathrm{Z}_\odot$. The abundance set adopted
for solar metallicity is given in table \ref{tab:solar}. The
individual abundances of most elements scales with the metallicity,
$X/H = Z \times (X/H)_\odot$. The two contrary elements are Helium and
Nitrogen. For Helium, the chemical yield from stars adds only a small
amount to the primordial abundance,
\begin{equation}\label{eqn:He/H}
\mathrm{He}/{\mathrm{H}} = 0.0737 +
0.0293\mathrm{Z}/{\mathrm{Z}_\odot}
\end{equation}

The nucleosynthetic status of Nitrogen is unusual in that it has both
primary and secondary nucleosynthetic components. The Nitrogen
abundance at different metallicities is given by;
 \begin{equation}\label{eqn:N/H}
\left[\mathrm{N}/\mathrm{H}\right] =\left[\mathrm{O}/\mathrm{H}\right]
\left(10^{-1.6} + 10^{\left(2.33 +
\log_{10}\left[\mathrm{O}/\mathrm{H}\right]\right)}\right).
\end{equation}
The origin of this and equation \ref{eqn:He/H}, as well as the
justification for the Solar
abundance set are discussed in Paper 1.

Because dust is present, the gas-phase abundances are depleted in
comparison to the total abundances. The depletion fractions adopted
are given in Table \ref{tab:depl}. In the new radiation pressure
dominated models the metals are depleted onto the dust, but for the
dust-free standard models this depletion is artificial, with the
metals effectively lost. This is done to ensure that the gas phase
abundances in both sets of models are the same, because the heavy
elements in the gas phase determine the cooling to a large extent and
also have some effect upon the ionization state within the nebula.

We use a simple power-law to represent the spectral energy
distribution (SED) of the ionizing source, with
\begin{equation}
F_\nu \propto \nu^\alpha \quad \nu_\mathrm{min} < \nu <
\nu_\mathrm{max}.
\end{equation}
and $\nu_\mathrm{min} = 5$eV and $\nu_\mathrm{max} = 1000$eV.  We
investigated four values of the power-law index $\alpha$, -1.2, -1.4,
-1.7 and -2.0. These indices encompass the `standard' values usually
adopted for modeling of the AGN spectrum. The factor of
proportionality, which determines the total radiative flux entering
the photoionized cloud is set by the ionization parameter at the front
of the cloud, $U_0 = S_\star/(n_0 c)$, where $S_\star$ is the entering
flux of ionizing photons, $n_0$ the initial density and $c$ the speed
of light.

This ionization parameter is the final parameter allowed to vary in
this family of models, with $\log U_0$ varying between 0.0 and -4.0,
sampled at intervals of -0.3, -0.6 and -1.0 dex. For the standard
isochoric models the initial density is a known quantity and hence the
ionization parameter at the front of the cloud is easy to
determine. In the dusty, isobaric models the initial density is not
known {\it a priori}. To estimate this value we assume a front end
temperature of $T_0 = 20,000$ K and obtain the density from the set
$P_0$. Depending upon the temperature reached at the front of the
model, the true density will vary, being overestimated at higher
temperatures and underestimated for cooler temperatures. This means
that the true ionization parameters in the upper range ($\log U_0 >
-2.0$) are actually larger than that given in the text, and are
smaller for the lower range ($\log U_0 < -2.0$). The general shape of
the curves in the line diagnostic diagrams is still correct. However,
the true range of ionization parameters is somewhat larger than
indicated on these diagrams.

%%%%%%%%%%%%%%%%%%%%%%%%%%%%%%%

\section{Diagnostic Diagrams\label{sec:LDDs}}

There are several considerations we must take into account in the choice of
line ratios to use in the line diagnostic diagrams (LDDs). These are:

\begin{enumerate}
\item The
lines must be strong enough to be measurable in most Seyferts.

\item The line diagnostic
diagram must provide physical insight  into the nature of the emitting
system.  For  example, the LDDs  put forward by  \citet{VO87} provided
clear diagnostics  of the excitation mechanism, with  NLRs, LINERs and
\hii\  galaxies  well separated.   Another  good  example  are the  UV
diagnostics of  \citet{ADT98} which  are able to  discriminate between
photoionization models, shock models, and shock+precursor models.

\item To minimize wavelength sensitive
effects  such  as  reddening  and  flux  calibration,  the  wavelength
separation between the lines that make up the ratio should be as small
as  possible. For  reddening this  becomes more  important  at shorter
wavelengths, especially around the  2200\AA\ extinction feature seen in
our galaxy.

\item The lines should be all observable using a common instrument and
the same  technology. For  example, a UV  line ratio versus  a Visible
line ratio may  be a good diagnostic, but it  is highly unlikely that
one could ensure the same  aperture for these observations. Hence line
ratio diagrams  should be made up  of ratios that lie  within a single
wavelength regime easily accessible by the same instrument.

\item Emission
lines  which are  usually  blended  with other  lines  must be  either
avoided or taken as a  blend, as deblending lines can greatly increase
the uncertainty of the measured flux.

\item One should avoid the use of lines which carry large flux errors
resulting from  the theoretical  models used.  An  example of  this is
[\oi]$\lambda6300$\AA\  which,  though  easily measurable,  is  highly
dependent upon the point at which the model has been truncated.

\end{enumerate}

%%%%%%%%%%%%%%%%%%%%%%%%%%%%%%%%%%%%%%%
%  Standard V + O diagrams
%%%%%%%%%%%%%%%%%%%%%%%%%%%%%%%%%%%%%%%
\subsection{Standard Optical Diagnostic Diagrams}\label{sec:VO}

The line diagnostic diagrams put forward by \citet{VO87} (V\&O) have
become standard diagnostics for emission line galaxies.  This is
because the diagrams not only consist of the some of the strongest
emission lines which are easily accessible to ground base telescopes,
but they also clearly distinguish between the excitation mechanisms of
the emission line galaxies.

Figure 1 shows the first of the V\&O diagrams;
[\nii]$\lambda6583/\mathrm{H}\alpha$ versus
[\oiii]$\lambda5007/\mathrm{H}\beta$.  Figure \ref{fig:VO}a
demonstrates the effect of different metallicities upon the dusty
models, with the Seyfert 2 observations from the original V\&O paper
given for comparison. The models plotted are for a density of 1000
\pccm\ and a power law index of $\alpha=-1.4$, and cover a range in
ionization parameter of $-4.0\le\log U_0\le0.0$. Lines of constant
ionization parameter are marked and labelled at set intervals.  Each
model curve represents a different metallicity, increasing from left
to right, with the metallicity of each curve labelled.  Also indicated
on the diagram is the effect of reddening upon the models. The
direction and magnitude of the arrow indicates the effect of an
external dust screen with 10 magnitudes of visual extinction using the
reddening curve from \citet{Calzetti01} (their equation 8).  The
diagram reveals the large spread in the [\nii]/H$\alpha$ ratio for the
set of photoionization curves, which is largely due to the varying
contributions of the secondary component of Nitrogen
nucleosynthesis. A comparison of the photoionization curves with the
observational data also reveals that most of the NLR in Seyfert
galaxies are characterized by super-solar metallicity, with a mean
value of about 2 Z$_{\odot}$. The sub-solar curves are unable to
reproduce the data.

Using this information, figure \ref{fig:VO}b compares both the dusty
models and the standard, dust-free isochoric models with the
observational data.  The different sets of models are labelled and
cover a range of power-law index from $\alpha=-1.2$ to $-2.0$.  Both
sets of models have a metallicity of 2 Z$_{\odot}$ and
n$_\mathrm{H}$=1000 \pccm, with each model curve ranging from $\log
U_0 =-3.0$ to $0.0$.  For both sets of models, as the power-law
becomes flatter (i.e.~harder spectral energy distribution), the
trajectories of the models are displaced from the bottom-left towards
the top-right region of the diagram.  The dusty models ``stagnate'' in
the region occupied by the observational data at high ionization
parameter. That is, as the ionization parameter increases, the dusty
models tend to return the same values for the ratios, and become
degenerate in terms of $U$ at the highest values. As for the dust-free
models, [\nii]/H$\alpha$ decreases continually with increasing
ionization parameter, as the overall ionization state of the gas
increases. This behavior provides a clear distinction between the
models and is one of the principal reasons successes of the dusty,
radiation pressure dominated photoionization models, as discussed in
\citet{DG02}.

There are several other points to note within the diagram. One is the
difference between the behavior of the two sets of models in the
[\oiii]/H$\beta$ ratio at low ionization parameter. At high ionization
parameter a difference is expected due to the effects of extinction by
dust and radiation pressure on dust, but at low ionization parameter
such effects should be minimal as dust is no longer the dominant
opacity. The difference between the models arises due to the
definition of the ionization parameter at the front of the cloud, as
discussed in \S \ref{sec:param}. In the case of $\log U_0$ of -3.0,
the dusty models overestimate the effective ionization parameter due
to the overestimate of the assumed initial temperature.  Hence the
dusty models display a weaker [\oiii]/H$\alpha$ ratio than the
dust-free set. The apparent difference is therefore an artifact of the
initial condition, and should be ignored. The lack of observational
points below $\log U_0 \sim -3$ is not a failing of the models, rather
it is due to the definition of Seyfert 2 galaxies. The region below
that occupied by the Seyferts is not empty, rather it contains the
class of galaxies classified as LINERs (see eg.~V\&O). LINERs may
include a subset which could be considered low power Seyferts, but
they also include shock excited emission line galaxies and possible
old starbursts as well, and thus we have not considered these within
our sample.

The other point to note is the spread in model sets due to the
variation in power-law index, $\alpha$.  For both sets of models a
flatter (harder) ionizing spectrum leads to an increase in both
ratios, especially [\oiii]/H$\beta$. This is due to the increase in
photons available to ionize these higher ionization species compared
to hydrogen.  At low ionization parameter, the region covered on the
diagram by the models of different power-law index is similar to that
found for models with different metallicity.  This indicates the
limited utility of these ratios as either a $Z$ or an $\alpha$
diagnostic. At high ionization parameter however, the variation of
$\alpha$ is clearly distinguished from that of $Z$ in the dusty
models, with the models at different $\alpha$ curving in upon each
other such that they deliver little spread in the [\oiii]/H$\beta$
ratio or the [\nii]/H$\alpha$ ratio.  The observational data is well
covered by the models, and the most reasonable values of $\alpha$
probably lie in the range $-1.4 < \alpha <-1.2$.

Figure \ref{fig:VO}c shows the V\&O diagram [\oi]$\lambda
6300$/H$\alpha$ versus [\oiii]$\lambda5007/\mathrm{H}\beta$, with the
data points being taken again from the same paper. Though the strength
of the [\oi]$\lambda 6300$ line may not be entirely trustworthy for
the reasons discussed previously, the line is strong and easily
measured and the ratio proves to be a good diagnostic.  As discussed
in Paper 1, our models are truncated at the point at which \hii/H
ratio drops below 1\%, such that the final temperature is a few 100
K. This means that the majority of the \oi\ emitting region is
encompassed within the model as the emission of \oi\ drops
below 1000K. Figure \ref{fig:VO}c demonstrates the effect of variation
of metallicity on both the dusty and the dust-free models. The models
have $U_0$ ranging from $10^{-3}$ to $10^0$, with $\alpha$ fixed at
$-1.4$ and the density at $1000$ \pccm.  The metallicity increases
from 0.25$Z_{\odot}$ for the leftmost curve to 4$Z_{\odot}$ on the
right. As both line ratios are ratios of oxygen to hydrogen lines, any
variations are due to changes in the temperature and ionization of the
gas.  The multi-valued nature of the curves arises due to the
combination of two effects.  As metallicity increases, the strength of
the metal emission lines initially increases due to the increase in
abundance.  However, increasing the metallicity also increases the
cooling efficiency of the gas and hence lowering the temperature.  At
high ionization parameter, there is a clear distinction between the
dusty and dust-free models, as seen in the previous diagrams. The
dusty model line ratios stagnate in a limited zone of parameter space,
while the dust free models display a large decrease in the
[\oi]/H$\alpha$ ratio with increasing ionization parameter. As in the
previous diagnostics, the observed data points are best fit by the
dusty models, with the curves at high ionization all clustered in the
region of interest. The data is best reproduced by the curves with a
metallicity of $Z=1-2 Z_{\odot}$, but, because oxygen is a primary
rather than secondary nucleosynthesis element, the distinction between
the curves of different metallicity is not as clear as in the case of
the [\nii]/H$\alpha$ ratio.

Figure \ref{fig:VO}d shows the effects of variation of $\alpha$ upon
the dusty model ratios. The models all have $Z=2Z_{\odot}$,
$n_\mathrm{H} = 1000$ \pccm, and each curve covers the range
$-3.0\le\log U_0\le0.0$. Each curve represents a different power-law
index, as labelled. This diagram more clearly demonstrates the spread
due to $\alpha$ at low metallicities, and the convergence of the
models in a narrow strip of parameter space, similar to that seen in
figure \ref{fig:VO}b. Though not shown here, the dust-free models also
appear very similar to the dust-free curves of figure \ref{fig:VO}b as
well. This diagram clearly indicates how well the dusty models
reproduce both the clustering and the absolute values of the
observational data. A value of $\alpha$ between $-1.4$ and $-1.9$
provides the best fit to the observed range of observational points.

The final V\&O diagram is [\sii]$\lambda\lambda6717,30$/H$\alpha$
versus [\oiii]$\lambda5007/\mathrm{H}\beta$.  Figure \ref{fig:VO}e
demonstrates the effects of metallicity variation upon the two sets of
models, while figure \ref{fig:VO}f demonstrates the effects of
power-law index variation upon both. Both figures use the same set of
fiducial parameters as before.

In most ways these diagrams are similar to the previous figures.  They
display a large spread due to metallicity and power-law index, and
show similar shapes to the previous model curves. At high ionization
parameter, the dusty and dust-free models separate as in the previous
figures, with the dusty models displaying the same stagnation in the
region of interest. The observational data is similarly best
reproduced by the dusty models, with a metallicity $Z\sim 1 Z_{\odot}$
and a power-law index $\alpha \sim -1.4 - -1.7$. Note that though the
$Z\sim 1 Z_{\odot}$ seems to better fit the data when metallicity
variation is considered, the diagram of the $\alpha$ variation (using
$Z = 2 Z_{\odot}$ to remain consistent) demonstrates that the
variation in power-law index may also be quite large.

Of the three V\&O diagnostic diagrams, the
[\nii]$\lambda5683/\mathrm{H}\alpha$ is the clearest metallicity
diagnostic thanks to the effects of the secondary nucleosynthetic
component of the nitrogen production. It has well separated
metallicity curves that cover a range which is much broader than that
due to power-law index variations. This means that it is not
degenerate with $\alpha$ as the other two diagrams are. This
diagnostic diagram shows that the mean metallicity in Seyfert NLRs is
$\sim 2 Z_{\odot}$ although it may truly range from $1 Z_{\odot}$ up
to $4 Z_{\odot}$. The result of a higher metallicity is not
surprising, as the new abundance set used and the depletion by dust
both act to reduce the gas abundance compared to previous models. A
value of $\sim 2 Z_{\odot}$ is approximately the same as the old Solar
values \citep{Anders89}, as has been estimated previously for NLR
metallicity.

An estimate of the average power-law index can be obtained by
consideration of all three diagrams. In these, the range
$-2.0\le\alpha\le-1.2$ encompasses all of the observed data, but more
typical values of the index range between $-1.4$ and $-1.7$.

The mean density of the Seyfert NLR clouds is harder to determine as
all three diagnostic diagrams show small variation due to density at
high ionization parameters when compared to the effects of the other
parameters, especially metallicity.

The self-consistency of the modelling is demonstrated by the fact that
the observed ratios on all three diagnostic diagrams are reasonably
reproduced by the same set of parameters, with a very broad range of
ionization parameters able to reproduce the tight clustering of the
data.

%%%%%%%%%%%%%%%%%%%%%%%%%%%%%%%%%%%%%%%%%%%%
% ADT UV diagrams + Inskip/Best
%%%%%%%%%%%%%%%%%%%%%%%%%%%%%%%%%%%%%%%%%%%%
\subsection{Standard UV Diagnostic Diagrams}

Whilst the optical diagnostic diagrams provide an excellent tool for
separating starburst galaxies from those with an AGN, these optical
line ratios are unable to conclusively distinguish between shocks and
photoionization as an excitation mechanism for active
galaxies. Ultraviolet emission lines, however, tend to be much
stronger in shocks than in simple photoionization models and hence
much better diagnostics for shock and photoionization separation. This
difference is due to the collisionally excited lines in the UV, like
\civ\ $\lambda 1549$, having increased emission at the higher
temperatures found in post-shock gas \citep[and references
therein]{do:su95,do:su96}. This temperature sensitivity also proves to
be useful in distinguishing between classical photoionization models
and the dusty models presented in this paper.

A further benefit of UV diagnostic diagrams is that in high redshift
active galaxies, the UV emission lines are shifted to the optical
band, and can be observed with ground-based telescopes. Hence the UV
diagnostic diagrams are better for determining the excitation
mechanism of the high-z galaxies than the standard optical ones which
may be difficult or impossible to observe.

UV diagnostic diagrams have been developed and explored in the case of
high-Z radio galaxies by \citet{VTC97} and comparisons of shock and
photoionization models have been made by \citet{ADT98} (hereafter
ADT). Here, we continue this exploration and comparison with the
inclusion of dusty photoionization models.

The first three UV diagrams of \citet{VTC97} use three UV line ratios;
\civ\ $\lambda 1549$/\ciii]$\lambda 1909$, \civ\ $\lambda 1549$/\heii
$\lambda 1640$ and \ciii]$\lambda 1909$/\heii $\lambda 1640$. The
observational data on each figure comes from ADT, with the stars
marking the high-$z$ radio galaxy (HZRG) observations and the
triangles representing HST observations of three nearby Seyferts NGC
1068, NGC 5643 and NGC 5728. In addition to the observational data
each figure is marked by an arrow indicating the size and direction of
the reddening vector by a foreground dusty screen with an $A_V$ of
three magnitudes. As for the previous section, the \citet{Calzetti01}
curve is used for reddening correction.

Figure \ref{fig:ADT} shows the effect of metallicity variation on the
\civ\ $\lambda 1549$/\ciii]$\lambda 1909$ versus \civ\ $\lambda
1549$/\heii\ $\lambda 1640$ diagnostic diagram. The models all have a
power-law index of $\alpha = -1.4$, and density $n_\mathrm{H} =
1000$\pccm. Each curve covers a range in ionization parameter of
$-2.3\le\log U_0\le0.0$, with each having a different metallicity as
labelled.  The range of ionization parameter shown here is smaller
than in the optical diagrams as \civ\ rapidly weakens for $U <
10^{-2}$. As in the optical diagrams, the line ratios stagnate at high
ionization parameter in the region where both the HZRGs and the
Seyfert galaxies lie. They also stagnate in metallicity as both line
ratios reach their maximum value at $Z \sim 1Z_{\odot}$. Any value
between $0.25 Z_{\odot} < Z <2.0 Z_{\odot}$ is equally effective in
reproducing the observations. A metallicity less than solar may even
be plausible for the HZRGs which are observed at early stages of
galaxy formation.

Following the format of the optical diagnostic diagrams, figure
\ref{fig:ADT}b demonstrates the effects of power-law index variation
upon both the dusty and dust-free models at a metallicity of
$2Z_\odot$, and $n_\mathrm{H}=1000$\pccm. The index becomes steeper
from top to bottom on the diagram for both sets of models and the
ionization parameter covers the same range as before. Little can be
inferred from this diagram other than that both the dusty and
dust-free models provide an equally good fit to the observational
points. However, it is significant that there are no observational
points above log(\civ/\ciii]) $>0.6$, just above the point where the
dusty models stagnate.

As discussed in Paper 1, care must be taken with these UV diagnostics
due to the presence of resonance lines like \civ. The radiative
transfer of these lines is not treated exactly within the
photoionization code, but rather approximated with an escape
probability, which includes the opacity effect of dust. This means
that there is a larger error associated with the ratios which involve
these lines and they should be treated as less certain.

The second UV diagnostic diagram is \civ\ $\lambda
1549$/\ciii]$\lambda 1909$ versus \ciii]$\lambda 1909$/\heii $\lambda
1640$, shown in figure \ref{fig:ADT}c, where the effect of metallicity
variation upon these line ratios in the dusty models is demonstrated.
Again all curves have $\alpha=-1.4$ and $n_\mathrm{H}=1000$\pccm\,
with $\log U_0$ ranging from $-3.0$ to $0.0$. The shape and form of
these curves are very similar to that found in figure
\ref{fig:ADT}a. Here the effect of metallicity variation is once again
more apparent at low ionization parameters, but both metallicity and
ionization parameter become degenerate at high ionization
parameter. All that can be said is that the dusty models fit the data
points in the range $\log U_0 >-2.5$ and $Z \lapprox 2.0 Z_{\odot}$.

In figure \ref{fig:ADT}d we assume a $2Z_\odot$ metallicity to compare
the dusty models with the dust-free models for these ratios. Shown in
this comparison is the effect of density variation on these two
models. The power-law index is $-1.4$ and ionization parameter is
$-3.0\le\log U_0\le0.0$ for these sets of models. In both sets of
models the density increases from $10^2$\pccm\ at the bottom to
$10^4$\pccm\ at the top. Clearly, although some density sensitivity is
present, the spread due to metallicity or spectral index is just as
large and hence these parameters cannot be separated by the use of
these ratios alone. Nonetheless, it is clear that the dusty
photoionization models provide a much better description of the range
of observations than do the dust-free models.

The third \citet{VTC97} diagram has \ciii]$\lambda 1909$/\heii
$\lambda 1640$ versus \civ\ $\lambda 1549$/\heii $\lambda 1640$.  As
before, we demonstrate first the effect of metallicity upon these
ratios for the dusty model, in figure \ref{fig:ADT}e, using
$\alpha=-1.4$, $n_\mathrm{H}=10^3$\pccm\ and $-3.0\le\log U_0
\le0.0$. Though rotated and flipped relative to the previous diagrams,
the curves on this diagram show a similar form to that seen on both
the other two diagrams.

Finally, figure \ref{fig:ADT}f shows the effect of $\alpha$ variation
with these ratios on both sets of models, using the standard
$Z=2Z_\odot$, $n_\mathrm{H}=1000$\pccm\ and $-3.0\le\log
U_0\le0.0$. The power-law index increases from left to right in both
sets of models.  This also has the same general form as the previous
diagrams. Once again, the line ratios of the dusty models stagnate in
$U_0$ and $Z$ in the region occupied by the observational
points. These dusty models provide a much better fit to the
observations than the dust-free models.

\citet{ADT98} also provided several other diagnostic diagrams, mainly
with the purpose of distinguishing shock excitation from
photoionization. These diagnostics also prove to be valuable in
further establishing the validity of the dusty, radiation pressure
dominated photoionization models.

Figure \ref{fig:ADTCC}a shows the sensitivity to metallicity and
ionization parameter of \civ\ $\lambda 1549$/\ciii]$\lambda 1909$
versus \cii]$\lambda 2326$/\ciii]$\lambda 1909$. The empty triangles
represent the Seyfert galaxies shown in the previous UV diagnostic
diagrams. This figure is somewhat more useful in distinguishing
between different metallicities, but unlike the previous three UV
diagnostics, it appears to indicate a metallicity greater than
$2Z_\odot$.  Another possibility is that these galaxies may also have
some shock component to them, which, as seen in figure 2d of ADT,
would bring the models closer to the observations.

This diagram is also somewhat sensitive to the power-law index, as
shown in figure \ref{fig:ADTCC}b, which displays both the dusty and
dust-free models are displayed with $2Z_\odot$,
$n_\mathrm{H}=10^3$\pccm\ and $-3.0\le \log U_0 \le0.0$. Clearly, the
dusty models once again provide a better description of the
observations than the dust-free models, which have too weak
\cii]/\ciii] ratio and too strong \civ/\cii] ratio at high ionization
parameter.

In figure \ref{fig:ADTNeO}a we examine an optical-near UV diagnostic,
[\nev]$\lambda 3426$/[\neiii]$\lambda3869$ versus [\oiii]$\lambda
5007$/H$\beta$. The effect of metallicity variation upon these ratios
in the dusty models is shown in figure \ref{fig:ADTNeO}a. Though the
spread is small compared to some of previous UV diagnostics, this
diagnostic has the additional benefit of the minimal error and
reddening in the [\oiii]/H$\beta$ ratio. Included on the diagram are
three observational data sets. The empty triangles represent the
Seyfert galaxies from the previous UV diagrams. The crosses are data
from four Seyfert 2 galaxies from \citet{Allen98}. The asterisks
represent observations of Seyfert 2 galaxies from the data set of
\citet{Koski78}. All three sets of data are consistent with a
metallicity somewhere between $1Z_\odot \le Z \le 4Z_\odot$. Although
the dusty models stagnate in terms of the ionization parameter in the
region occupied by the data, not as much can be inferred from this
because the dust-free models are multi-valued in $U_0$ in this region
as well.  This is shown in figure \ref{fig:ADTNeO}b, which plots
curves of different $\alpha$ for the two models at $Z=2Z_\odot$,
n$_H=1000$\pccm and $-2.6 \le \log U_0 \le 0.0$. $\alpha$ increases
with increasing [\oiii]/H$\beta$. The two sets of models differ in
their behavior in $\alpha$, with the dusty model curves stagnating in
a very restricted region of parameter space but the dust-free models
becoming more widely separated. The fact that the stagnation point of
the dusty models agrees with the observational points gives further
credence to the general validity of this class of models.

In addition to the ADT paper there have been more recent
investigations of UV diagnostics, such as \citet{Inskip02a,Inskip02b}
who used the diagnostics to examine high redshift ($z\sim 1$) radio
galaxies. The redshift of these galaxies brought several of the UV
diagnostic lines into the optical range, making them accessible to
ground based telescopes, with long enough integration times.  Due to
the fact that these galaxies have very energetic expanding radio
lobes, they are expected to be excited by both shocks and by nuclear
photoionization. Being able to distinguish between these differing
excitation mechanisms is important in our understanding of these
high-$z$ AGN.

The UV diagnostic diagram chosen by \citet{Best00b}, and expanded by
\citet{Inskip02a}, combines the same neon line ratio of the previous
diagram; [\neiii]$\lambda3869$/[\nev]$\lambda 3426$ but plotted
against the UV carbon line ratio \ciii]$\lambda 1909$/\cii]$\lambda
2326$. This diagram is very good for distinguishing between shock
excitation and photoionization. It also turns out to be very good for
distinguishing between the dusty models and the dust-free models
(figure \ref{fig:Inskip}a). The HZRG observations from
\citet{Inskip02a} (their figure 14) are marked on the
diagram. Triangles indicate the 6C radio galaxies and stars the 3C
radio galaxies. The ionization parameter is restricted to the range
$-2.3\le \log U_0 \le 0.0$ because [\nev] is a very high ionization
species and becomes very weak at lower ionization parameters. Figure
\ref{fig:Inskip}a shows the sensitivity of the emission line ratios to
the metallicity. Again, many of these radio galaxies sit in the region
of the diagram that is consistent with high ionization parameter;
$\log U_0 \gapprox -1.0$. However, little can be inferred about the
metallicity. The radio galaxies with weak \ciii]/\cii] ratios (which
are believed to be shock or shock+precursor excited
\citep{Best00a,Best00b,Inskip02a, Inskip02b}) cannot be reproduced by
these dusty models without resorting to very high ($Z>4Z_\odot$)
metallicities. This is similar to figure \ref{fig:ADTCC}.

Figure \ref{fig:Inskip}b demonstrates the use of this diagram in
distinguishing between dusty and dust-free photoionization models. The
model curves, with $Z=2Z_\odot$, $n_\mathrm{H}=1000$\pccm, vary in
power-law index, ranging from -1.2 at low \ciii]/\cii] to -2.0 for the
models with high \ciii]/\cii]. The dusty models are clearly able to
reproduce the HZRG observations that are believed to be photoionized,
whereas the dust-free models would imply a much flatter power-law
($\alpha < -1.2$) or else a higher metallicity to even approach the
observational points.

Though our ability to derive a clear definitive set of parameters for
the Seyfert Galaxies and the high redshift radio galaxies is limited
in these UV diagnostic diagrams due to the degenerate nature of many
of the curves, one detail stands out in all diagnostics; the dusty,
radiation pressure dominated models provide undeniably the better fit
to the observations over the standard dust-free models. In all cases
the dusty models not only reproduce the data well, but also tend to
become degenerate in terms of the ionization parameter precisely in
the region occupied by the observations.

%%%%%%%%%%%%%%%%%%%%%%%%%%%%%%%%%%%%%%%
% New UV, Vis and IR diagrams - especially metallicity diagnostics
%%%%%%%%%%%%%%%%%%%%%%%%%%%%%%%%%%%%%%%
\subsection{Further Useful Diagnostic Diagrams}

The previous sections demonstrated the utility of the dusty, radiation
pressure dominated models. We will now explore several new or
relatively unexplored line ratio diagrams, not only to gain a deeper
understanding of the dusty photoionization models but to also find
diagrams which can provide diagnostics of the metallicity, density and
the slope of the ionizing power-law.

These diagrams extend the wavelength base from the far-UV into the
near and mid-IR. They adhere to the guidelines of \citet{VO87}
re-iterated at the start of section \ref{sec:LDDs}. The program used
to create these diagnostics is available at\\
\verb|http://www.mso.anu.edu.au/~bgroves/linedata|. Also available
there is the model data used to create these diagnostics, which were
discussed in depth in Paper 1. This includes the standard, dust-free,
undepleted models discussed in Paper 1, if the readers wish to examine
the diagnostic diagrams of this model.

For the dusty models we adopt a fiducial model which has a metallicity
of $2Z_\odot$, a power-law index of $\alpha=-1.4$ and a density of
$n_\mathrm{H}=1000$\pccm\ following the results of the previous
sections. Each of these parameters is adjusted separately to determine
the sensitivity of each line diagnostic to each parameter in turn. For
the ionization parameter, the greatest possible range is
explored. This turns out to be quite restricted for the line ratios
which involve high ionization species.

%%%%%%%%%%%%%%%%%%%%%%%%%%%%%%%%%%%%%%
\subsubsection{UV Diagnostics}

The benefits of UV diagnostics in providing insight into the physics
of high redshift active galaxies and in distinguishing between
photoionization and shock excitation has been examined in depth in
earlier works by \citet{VTC97,ADT98,Best00a,Best00b} and
\citet{Inskip02a,Inskip02b}. However, with the deeper surveys of high
redshift galaxies currently underway, we can expect that diagnostic
plots involving lines in the far-UV will soon become even more useful.

The first far-UV diagnostic takes two line ratios from ADT; \ciii\
$\lambda977$/\ciii]$\lambda1909$ and
\niii$\lambda991$/\niii]$\lambda1750$. Both ratios are temperature
sensitive and hence are good at separating shock excitation from
photoionization (see figure 3 of ADT). When plotted against a ratio
which increases steadily with $U_0$, such as \civ\ $\lambda
1549$/\ciii]$\lambda 1909$ they also provide reasonably good
metallicity diagnostics, although the spread due to $\alpha$ or
n$_\mathrm{H}$ are similar in range. Both of these line ratios have a
similar dependence on the electron temperature. Thus, when plotted
together, variations in the metallicity, power-law index and
ionization parameter are relatively indistinguishable. However, the
ratios do provide a diagnostic which is able to distinguish variations
in the density, as shown in figure \ref{fig:UV1}.  This arises mainly
through the nitrogen ratio. The \niii]$\lambda1750$ line can be
represented as a five level system, containing the lines that make up
the 1750 line and a fine structure line at 57 \mum. As the density
increases, the timescale for collision from the fine structure level
becomes less than the emission timescale, and the flux that would
normally be emitted through 57 \mum, is emitted at 1750 \AA. As the
\ciii\ $\lambda977$ resonance line is reasonably unaffected by
density, this ratio becomes density sensitive. The density sensitivity
is reversed at low $U$ due to the lower temperature and hence lower
collisional excitation rate.

One problem with these ratios though is that the reddening is quite
large due to the short wavelengths of the lines and the large
wavelength separation between numerator and denominator. Corrections
due to reddening are also quite uncertain at these short wavelengths,
hence the absence of the reddening arrow on figure \ref{fig:UV1}. The
small ($\sim 0.1$ dex) separation between the three densities and the
likely large absolute flux errors resulting from the uncertain
reddening corrections means that this diagram is probably not useful
as a density diagnostic.

In figure \ref{fig:UV2} we plot \civ\ $\lambda1549$/\ciii]$\lambda
1909$ versus \ciii$\lambda977$/\niii$\lambda991$. These line ratios
are much less sensitive to both temperature and reddening than those
used in the previous diagram. The sensitivity to $\alpha$ is also
minimized as a consequence of the insensitivity to temperature. The
ratios also show very little variation due to density changes. However
the \ciii/\niii\ ratio is quite sensitive to changes in
metallicity. This is thanks to the secondary nucleosynthetic component
of the nitrogen abundance. The \civ/\ciii] ratio is primarily useful
in distinguishing between different values of $U_0$. In addition to
the effects of metallicity variations, what is also visible is the
stagnation at high values of $U_0$ due to dust and the radiation
pressure upon it.

Figure \ref{fig:UV3} exploits the metallicity dependence further by
plotting \ciii]$\lambda1909$/\niii]$\lambda1750$ versus
\ciii$\lambda977$/\niii$\lambda991$. Whereas the previous diagram only
separated metallicity on one axis, here both axes are ratios of C/N
and show large separation between metallicity curves.  In addition,
the reddening corrections are small, since both pairs of lines are
closely separated in wavelength. Thus we have a reasonably sensitive
metallicity diagnostic. The only failing of this diagnostic is the
short wavelengths of the \ciii/\niii\ ratio which may make this
diagnostic difficult to observe.

Our final far-UV diagnostic diagram also provides a metallicity
diagnostic, plotting \ovi $\lambda\lambda1032,8$/\civ $\lambda1549$
versus \nv $\lambda1240$/\civ $\lambda1549$ (figure
\ref{fig:UV4}). This diagnostic diagram should prove to be useful in
several ways. Firstly, it involves only strong lines and is thus
easily measurable.  Secondly, the response of the curves to $\alpha$
and $n_\mathrm{H}$ variations is negligible compared to the
metallicity variations. Thirdly, the sensitivity of the \ovi/\civ\ to
ionization parameter is strong. Thus this diagram provides an
excellent metallicity and ionization parameter diagnostic for
photoionized narrow line clouds.

%
% Add something about OVI doublet being a problem??
% 
%%%%%%%%%%%%%%%%%%%%%%%%%%%%%%%%%%%%%%
\subsubsection{Optical Diagnostic Diagrams}

The optical spectral range is easy to observe and provides a large
number of lines from which to form diagnostics. Many of these line
ratio diagnostics are either temperature or density sensitive as well.

Figure \ref{fig:NeO} plots the near-UV ratio
[\nev]$\lambda3426$/[\neiii]$\lambda 3869$ against the density
sensitive ratio [\oii]$\lambda\lambda3727,9$/[\oiii]$\lambda5007$. We
show both the density and ionization parameter dependence of both the
dusty and dust-free models. The density increases with decreasing
[\oii]/[\oiii] ratio, from $10^2$ \pccm\ at the top to $10^4$ \pccm\
at the bottom. Included on this diagram are the observations of
Seyfert 2 NLRs from \citet{Koski78} (asterisks) and \citet{Allen98}
(crosses). Note that some of the Koski sample has been removed due to
contamination by starbursts \citep[see, eg][]{Kewley01} Within this
diagram the oxygen ratio provides sensitivity to density variations,
while the neon ratio provides sensitivity to ionization parameter.
Though density variations generate a strong response in the
[\oii]/[\oiii] ratio, changes in the other parameters also have some
consequences. Variation in the power-law index affects the
[\nev]/[\neii] ratio (as seen in figure \ref{fig:Inskip}a) and may be
confused with the effects of density and ionization parameter.  In
terms of metallicity changes, the only major effect is seen at high
$U_0$ in the dusty models, where it creates a similar spread in the
[\oii]/[\oiii] ratio to density.  So, overall, this diagram provides a
reasonable density diagnostic.  What is obvious from figure
\ref{fig:NeO} is that this diagram is able to strongly distinguish
between the two sets of photoionization models. The large separation
at high ionization parameter clearly demonstrates the success of the
dusty model in reproducing the observations. This diagnostic ability
is largely due to the [\oii]/[\oiii] ratio and arises because of the
differences in density structure of the isochoric, dust-free model and
the isobaric, dusty model.

The use of this ratio as a model diagnostic is explored further in the
next diagram, figure \ref{fig:OIIOIII}, which plots
[\oii]$\lambda\lambda3727,9$/[\oiii]$\lambda5007$ against
[\oiii]$\lambda5007$/H$\beta$. This diagram was one of the original
diagnostic diagrams suggested by \citet{BPT81} and has also been used
by \citet{bws96} to explore the $A_{M/I}$ model.  In our figure both
the dusty and dust-free models have been plotted, displaying curves of
different power-law index. In both cases a flatter power-law
corresponds to an increase in the [\oiii]/H$\beta$ index.  The
observations of \citet{Koski78} and \citet{Allen98} are displayed
using the same symbols as the previous diagram. In this diagnostic,
the [\oiii]/H$\beta$ provides a standard diagnostic ratio which helps
to separate the two models, as well as distinguish the different
$\alpha$ curves.  The applicability of the [\oii]/[\oiii] ratio as a
model diagnostic is clear in this figure, with the two model sets
separating above $\log U_0 =-2.0$. The dusty model curves
characteristically stagnate in ionization parameter above this value,
becoming degenerate in $U_0$ in the region occupied by the
observations. The dust-free models continue past this region to
proceed to smaller values of the [\oii]/[\oiii] ratio.

In the next diagram, figure \ref{fig:HeO}, we use the [\oii]/[\oiii]
ratio to introduce another model diagnostic ratio, \heii $\lambda
4686$/H$\beta$. This diagnostic diagram was first introduced by
\citet{bws96} to distinguish the $A_{M/I}$ model, as it was the
failings of the standard photoionization model to reproduce the
observed NLR \heii/H$\beta$ ratios that led to the development of this
model.  The \heii/H$\beta$ ratio is distinct in the diagnostic ratios
examined so far in that it is the dust-free models which stagnate at
high ionization parameter, not the dusty models.  In figure
\ref{fig:HeO} we display curves of different $\alpha$ for both of
these models, with $\alpha$ increasing from right to left.  The
stagnation of dust-free model in the \heii/H$\beta$ ratio at high
ionization parameter can be seen in the way the two model sets
diverge. The way in which the dusty model stagnates in the
[\oii]/[\oiii] ratio at high $U_0$, while still heading towards larger
values of the \heii/H$\beta$ ratio, reproduces the observations
extremely well, especially in comparison to the inverse behavior seen
in the dust-free models. The success of the dusty model in attaining
the high \heii/H$\beta$ ratio with the observed [\oii]/[\oiii] values
comes from its ability to maintain both a high and low-ionization zone
at high ionization parameter. This is similar to the main idea of the
$A_{M/I}$ model, except it is dust which provides the absorbed
spectrum not "matter bound clouds". The absorption by dust further
assists by preferentially removing the H ionizing photons to the He
ionizing ones, and hence increasing the \heii/H$\beta$ ratio.

Figure \ref{fig:HeO} is also able to distinguish $\alpha$, with a
large spread in the curves across the \heii/H$\beta$ ratio, especially
in the dust-free models. Density variations cause similar effects to
that seen in figure \ref{fig:NeO}, with the [\oii]/[\oiii] ratio
providing most of the sensitivity. The influence of metallicity is
interesting, as it appears to be negligible in the dust-free models,
yet the dominant parameter in the dusty models at high ionization
parameter.

The consequences of metallicity variations upon the \heii/H$\beta$
ratio are better seen in figure \ref{fig:Evansa}, which consists of
\heii\ $\lambda 4686$/H$\beta$ versus the temperature sensitive ratio
[\oiii]$\lambda 4363$/[\oiii]$\lambda 5007$ ($R_{OIII}$). The
\citet{Koski78} and \citet{Allen98} observations are included for
comparison.  On the diagram we display the effects of metallicity
variation upon the dusty models. The effect of metallicity variation
is predominantly seen in the $R_{OIII}$ ratio. An increase in the
metallicity results in a decrease in the nebula temperature and thus a
decrease in $R_{OIII}$, as metals dominate the cooling processes in
nebulae. These affects are discussed in detail in Paper 1. In the
\heii/H$\beta$ ratio the effects of metallicity are not obvious until
$U_0 > 10^{-2}$. Although the change in abundance affects the \heii\
emission, the change in emission with metallicity at high $U_0$ is
caused by the change in temperature and the increase in dust.  As
shown in \S 5.1 of Paper 1, an increase in metallicity leads to drop
in temperature in the \heii\ zone relative to the H$\beta$ emitting
region, and hence an increase in the recombination to \heii\ relative
to H. However this effect only contributes $\sim$10\% to the
spread. The major effect is due to the increase in the dust to gas
ratio because of the increase in metal abundance. This increases the
total dust opacity, which preferentially removes the H ionizing
photons to the He ones (see figure 2 in Paper 1), thus increases the
\heii\ emitting column relative to the \hii\ column and hence the
\heii/H$\beta$ ratio.  The spread in model curves is not seen at low
ionization parameter because the dust is no longer the dominant
opacity and hence does not control the emitting column.

The [\oiii]$\lambda 4363$/[\oiii]$\lambda 5007$ ($R_{OIII}$) ratio is
actually one of the standard line ratios, used because of its strength
and temperature sensitivity. Figure \ref{fig:ROIII} shows a standard
diagnostic diagram plotting $R_{OIII}$ against [\oiii]$\lambda
5007$/H$\beta$. Shown are the effects of density variation on both the
dusty and dust-free models, with density increasing from left to right
at $\log U_0=-3.0$. Included with the usual \citet{Koski78} (crosses)
and \citet{Allen98} (asterisks) observations, are data from
\citet{Tadhunter89} (triangles), who specifically looked at the
``Temperature problem''; the inability of photoionization models to
reproduce the high $R_{OIII}$ ratio. This failure is easily seen in
the way the dust-free model curves wrap around at high $U_0$, not
reaching the high $R_{OIII}$ region occupied by the data. What can
also be seen in the figure is the success of the dusty models in
attaining $R_{OIII}$ values similar to those observed. This success is
due to several reasons, the main being the hardening of the spectrum
by dust absorption and the contribution of photoelectric heating by
dust to the temperature.

Combining two of the previous ratios, the \heii/H$\beta$ versus
$R_{OIII}$ diagram is one of the more powerful diagnostic diagrams. As
shown in figure 10 in \citet{Evans99}, it clearly separates the
different possible excitation mechanisms. It is also a diagram in
which the standard photoionization model conspicuously fails to
reproduce the observations, being too cool to get strong $R_{OIII}$
and unable to attain the strong \heii/H$\beta$. Figure
\ref{fig:Evansb} demonstrates this failure, along with the success of
the dusty, radiation pressure dominated models in reproducing strong
\heii/H$\beta$. Curves of different density have been plotted for both
models sets, with density increasing with increasing $R_{OIII}$. There
is a clear separation between the dusty and dust-free models, with the
dusty models turning over in both ratios before the observations are
reached. In this instance, the dusty models have too low a $R_{OIII}$
ratio to successfully reproduce the observations. A flatter power-law
and lower metallicity can better reproduce the observations due to the
increase in temperature and the increase in \heii\ ionizing photons,
but such changes disagree with the previous diagnostic diagrams. The
data itself may also be in error due to the possibility of
overestimating such a weak line like [\oiii]$\lambda 4363$ (especially
in comparison with a strong one like $\lambda 5007$). Either way this
diagram indicates that the models still require some further
examination. The dust alleviates part of the temperature problem, but
some additional heating source is still needed, such as small shocks
or turbulent heating.

The \heii/H$\beta$ diagnostic series continues with \heii/H$\beta$
versus [\oiii]$\lambda 5007$/H$\beta$ in figure \ref{fig:HeOIII}. This
use of the standard diagnostic ratio provides a clear distinction
between the two models, plotted here with density the varying
parameter. In both models the higher density corresponds to a stronger
[\oiii]/H$\beta$ ratio. This diagram is a simple and clear
justification of the dusty model over the dust-free one.

The final diagnostic diagram in the \heii/H$\beta$ series is against
[\nii]$\lambda 6583$/H$\alpha$ in figure \ref{fig:HeNII}. In
\ref{fig:HeNII}a we show dusty model curves of different metallicity,
along with the \citet{Koski78} and \citet{Allen98} data. The use of
[\nii]/H$\alpha$ as a metallicity diagnostic has been discussed
previously but here the \heii/H$\beta$ provides a clear ionization
parameter diagnostic. This enables a better determination of the
metallicity, as well as $U_0$.

As a model diagnostic this diagram is also very good, as demonstrated
by figure \ref{fig:HeNII}b. The variation due to density in both
models is negligible when compared to the separation of the dusty and
dust-free model curves at high ionization parameter ($U_0
>10^{-2}$). Even the dispersion due to metallicity variations provides
only small confusion in distinguishing the two photoionization models.
At low ionization parameter ($U_0 < 10^{-2}$), the models become
indistinguishable as the effects of dust and radiation pressure on
dust are relatively small below this value. The data from
\citet{Koski78} and \citet{Allen98} again agree with the dusty model.

The next optical diagnostic diagram examines the partially ionized
zone of the NLR clouds described in Paper 1 by analyzing the low
ionization states of oxygen and nitrogen. Figure \ref{fig:NO} plots
[\ni]$\lambda 5200$/[\oii]$\lambda\lambda7318,24$ versus
[\oi]$\lambda6300$/[\nii]$\lambda6583$ for both sets of models, with
density decreasing from left to right for both model sets.  As
mentioned before, the low ionization species like \oi\ and \ni\ are
somewhat untrustworthy in the models due to their dependence upon the
model termination point, which is at the point where H is less than
1\% ionized in our models. However the twisting nature of these
diagnostic curves make this diagram worth investigating.
 As expected, for $\log U_0 < -2.0$ the models are very
similar, but above this value they diverge. The dust-free model curves
turn back upon themselves, heading towards small values of the
[\ni]/[\oii] and [\oi]/[\nii] ratios. The dusty models continue to
increase in both ratios, largely because the partially ionized region
continues to grow in the dusty models with $U_0$, as the dust opacity
acts to harden the ionizing spectrum.

The next diagram (figure \ref{fig:NIISII}) definitively demonstrates
the effectiveness of the dusty models. The ratios
[\sii]$\lambda\lambda6717,30$/H$\alpha$ versus
[\nii]$\lambda6583$/H$\alpha$ are plotted for the dusty models,
displaying the curves of different metallicity. These two ratios
together effectively display what was seen with the \citet{VO87}
curves. When the ionization parameter ranges over $-4.0 \le \log U_0
\le 0.0$ the dust-free models cover a range of 3 dex in both ratios
(as can be seen in figure 1). For the dusty, radiation pressure
dominated models however, over this large range in ionization
parameter each metallicity curve only covers a range of $\sim$0.5 dex
in both ratios. Even more indicative is that this region of stagnation
is the same region as occupied by the observations.

The final optical diagnostic diagram considers the combined ratio
[\oii]$\lambda\lambda3727,9 \times$[\sii]$\lambda\lambda6717,30$
/[\sii]$\lambda\lambda4067,76 \times$[\oii]$\lambda\lambda7318,24$. In
figure \ref{fig:OIISII} the variations in density for the dusty models
are plotted on this ratio versus the ionization parameter diagnostic
 [\oiii]$\lambda 5007$/H$\beta$. The [\oii][\sii]/[\sii][\oii] ratio
 minimizes the
effects of variations in both metallicity and power-law index, as the
sensitivity of each ratio is cancelled out by its inverse. The ratio
has both sulfur and oxygen, and short and long wavelength emission
lines on both numerator and denominator, leaving a ratio which
experiences little influence from abundance variations and reddening.
What it does leaves is a ratio strongly dependent upon density,
clearly demonstrated in figure \ref{fig:OIISII}.

Though the previous diagrams are only a small sample of the possible
optical diagnostic diagrams, they provide enough diagnostics such that
they can distinguish the excitation mechanism and verify that the
dusty, radiation pressure dominated model is the correct paradigm for
photoionization. Together they can also provide estimates for the
parameters that define the nebulae; density, metallicity and ionizing
spectral energy distribution.

%%%%%%%%%%%%%%%%%%%%%%%%%%%%%%%%%%%%%%
\subsubsection{Near-Infrared Diagnostic Diagrams}

Though the Near-Infrared experience little extinction due to dust,
there is a scarcity of strong lines in this spectral region and thus
limited choices for diagnostics. The noble elements, Ne and Ar
dominate this spectral region and provide good diagnostic emission
lines as they are not depleted onto dust.

The first diagnostic, figure \ref{fig:IR1},\\ plots [\nevi]$\lambda
7.652$\mum/[\neii]$\lambda 12.8$\mum\ versus
 [\arvi]$\lambda4.53$\mum/[\arii]$\lambda6.98$\mum\ and displays the
effects of variation of $\alpha$ upon the dusty model. Note that as
they both contain emission lines from high ionization species, the
ratios from the model rapidly disappear as the ionization parameter
becomes small ($U_0<10^{-2}$). The IR diagnostic diagram is similar to
both the UV and optical diagrams in that it demonstrates the
stagnation of the dusty models at high ionization parameter. It also
shows the same divergence of the dusty and dust-free models above
$\log U_0 =-2.0$, though not indicated here. This particular
diagnostic diagram is interesting in that it displays a strong
sensitivity to variations in $\alpha$, with both the influence of
metallicity and density being negligible. The sensitivity to
metallicity is removed through having ratios of the same element,
making this diagram a very good diagnostic of the ionizing SED. The
sensitivity to $\alpha$ arises from the difference in ionization
states of the numerator and denominator in both ratios.

The next diagram also displays a similar sole sensitivity to
$\alpha$.\\ Figure \ref{fig:IR2} plots [\nevi]$\lambda
7.652$\mum/[\neii]$\lambda 12.8$\mum\ versus [\ariii]$\lambda
8.98$\mum/[\arii]$\lambda 6.98$\mum\ displaying curves of different
power-law index for both the dusty and dust-free models. Apparent in
the diagram is the divergence of the two model sets at large values of
ionization parameter ($U_0>10^{-2}$). Also conspicuous is the
characteristic stagnation in the dusty models at these high values of
ionization parameter. The power-law index of each curve decreases with
increasing value of the Argon ratio for both sets of models, with
$-2.0$ at the top to $-1.2$ for the bottom curve. The use of this
diagram as an $\alpha$ diagnostic is clear, and the negligible
reddening in this spectral region make this an ideal diagnostic for
heavily obscured NLRs. Note that the Ne ratio has a large separation
at low $U_0$, which is not obvious due to the different scales on each
axis. This is as when the power law steepens, it drops the number of
\nev\ ionizing photons below a critical point and \nevi\ becomes very
weak.

With these two diagrams we have what appears to be rare in the UV and
optical; a pure power-law diagnostic. Thus, if an active galaxies
emission lines are measured for all three spectral regions, the use of
line diagnostic diagrams can not only put strong constraints on the
excitation mechanism but also the density, metallicity and ionizing
spectrum of the emission line region.

%%%%%%%%%%%%%%%%%%%%%%%%%%%%%%%%%%%%%%%%%%

\section{Discussion}

Through the diagnostic diagrams and comparisons with observations in
the previous section we have demonstrated the validity of the dusty,
radiation pressure dominated models as the paradigm for
photoionization. The ability to reproduce both the observations and
observed clustering of narrow line regions with similar parameters in
diagnostic diagrams is strong evidence for the model. However, this
work does not negate all work based upon the standard photoionization
model. The active galaxies which were believed to be photoionized
still appear to be so.  The parameters estimated for these regions by
the standard model are in general still correct. What this new
paradigm does provide is a better diagnostic for the excitation
mechanism, an explanation of how and where the standard model failed
and better diagnostics to obtain more accurate estimates for the
parameters that define these regions.

Neither does this new paradigm prove wrong the previous improvements
upon the standard model, such as the $A_{M/I}$ model \citep{bws96,
bwrs97} and the ``locally optimal emitting cloud'' model
\citep{Baldwin95, Ferg97}. Rather, this work, itself an improvement
upon the standard model, provides a self consistent and self-contained
physical basis for the emission of both high- and low-ionization lines
without the necessary combining of clouds of different densities and
ionization states.

As such, the dusty, radiation pressure dominated models can be
considered as an explanation for the assumptions of the previous
models.  For example, the structure of the clouds in the dusty model,
a depiction of which is displayed in figure 1 of \citet{DG02}, closely
matches one of the geometrical distributions suggested for the
$A_{M/I}$ models \citep[][figure 4b]{bws96}. The dusty model provides
both the low ionization, ionization bounded component and the high
ionization, matter bounded component within a single model, as shown
in the ionization structure diagrams given in Paper 1. It does this
through self-shielding, such that the high-ionization component sees
the direct spectrum while the low ionization component near the
ionization front sees the self absorbed spectrum. The dusty models
also provide an explanation for the pressure difference between the
matter bounded and ionization bounded components of the $A_{M/I}$
model through the effects of radiation pressure on dust.  The final
supporting point is that the dusty, radiation pressure dominated
models also answers the problem emission line ratios that the
$A_{M/I}$ models was originally created to solve. As shown in figure
\ref{fig:Evansa} and \ref{fig:Evansb}, the dusty model is able to
produce strong \heii/H$\beta$ and strong [\oiii]$\lambda
4363$/[\oiii]$\lambda 5007$ ($R_{OIII}$), through the combination of
ionization zones and dust photoelectric heating. The main distinction
between the results of the dusty models and the $A_{M/I}$ models is
also the major improvement; the stagnation in ionization parameter
characteristic to the dusty model.

Just as the dusty, radiation pressure model does not prove incorrect
the previous photoionization models, neither does it prove wrong the
shock excitation models. The shock models and the photoionizing
shock+precursor models require some form of mechanical energy
input. In the absence of such an input the emission line clouds must
be photoionized. However, if a jet or strong outflow is present within
the AGN then shock excitation will play some part. In most cases the
contribution from shock excited clouds is either indistinguishable or
negligible compared the contribution from photoionized clouds. However
in some cases, the emission from the shock excited clouds can become
noticeable or even dominate the emission from photoionized
clouds. Such a case is seen in the UV diagnostic shown in figure 5
when compared to the shock models and observations from figure 14 in
\citet{Inskip02a}. Most of the radio galaxies are reproduced quite
well by the dusty models. However a few of the observed high-$z$ radio
galaxies have a lower \ciii]/\cii] ratio then is obtainable for the
values of [\neii]/[\nev] seen. However, as discussed in
\citet{Inskip02a}, these outlying galaxies can either be easily fit by
the shock+precursor model or a combination of shock+precursor and
dusty model. \citet{Inskip02a} also suggest that the balance between
the two components depends upon the radio source size and hence jet
structure, which may indicate some form of jet-cloud interaction. Even
so, these observations demonstrate that more needs to be understood
about how these two methods of excitation can coexist and how they
interact.

%%%%%%%%%%%%%%%%%%%%%%%%%%%%%%%%%%%%%%%%%%

\section{Conclusion}

First introduced in \citet{DG02} the dusty, radiation pressure
dominated photoionization model provides a self-consistent explanation
for the emission from the narrow line regions of AGN. Within this work
we have continued the examination of this new paradigm begun in
\citet{GDS03}.

Through the comparison of observations on standard optical line
diagnostic diagrams of \citet{VO87} and the UV diagnostics of
\citet{ADT98} we have clearly demonstrated the validity of the dusty
model as the new paradigm for photoionization. The dusty model,
through the stagnation of the ionization parameter at large values,
provides a simple explanation for the small variation of observed
Seyfert NLR ratios. This stagnation is due to the effects of radiation
pressure upon dust and is characteristic to these models. The
significant point is that the dusty model is able to do this over both
optical and UV ratios, without depending upon large variations in
other parameters such as density or metallicity.

As well as verifying the dusty model,we have also explored the effects
that the variation of density, metallicity and ionizing spectrum have
upon both the new dusty paradigm and the standard photoionization
models. With this exploration, we have demonstrated that several known
line ratio diagrams can be possibly used as diagnostics of these three
parameters.

In addition to the previously explored diagnostics, we have introduced
several new line diagnostic diagrams, covering UV, optical and IR
ratios. These diagnostics provide further tests for the dusty model as
well as providing diagnostics for metallicity, density, ionizing
spectrum and ionization parameter.

These results not only provide an explanation for what has not been a
fully understood observation for years but also provide ways in which
to understand further the processes involved in the NLR and extended
NLR of AGN.

\begin{acknowledgements}
M.D.~acknowledges the support of the Australian National University
and of the Australian Research Council through his ARC Australian
Federation Fellowship and M.D.~and R.S.~acknowledges support through
the ARC Discovery project DP0208445. The authors would like to thank
the referee for their incisive and detailed comments which have much
improved these papers.
\end{acknowledgements}

%%%%%%%%%%%%%%%%%%%%%%%%%%%%%%%

\clearpage
%
%  Tables
%
\begin{table}
\begin{center}
\caption{Solar Abundance Set\label{tab:solar}}
~\\
\begin{tabular}{|lc|lc|}

\tableline
Element & Abundance\tablenotemark{a} & Element &
Abundance\tablenotemark{a} \\
\tableline
\tableline
H  & ~0.000 & Al & -5.51~ \\
He & -0.987 & Si & -4.49~ \\
C  & -3.61~ & S  & -4.80~ \\
N  & -4.20~ & Cl & -6.72~ \\
O  & -3.31~ & Ar & -5.60~ \\
Ne & -3.92~ & Ca & -5.65~ \\
Na & -5.68~ & Fe & -4.54~ \\
Mg & -4.42~ & Ni & -5.75~ \\
\tableline
\end{tabular}
\end{center}
%\enddata
\tablenotetext{a}{All abundances are logarithmic with respect to Hydrogen}
\end{table}
\begin{deluxetable}{lr}
\tablecaption{Depletion Factors
\label{tab:depl}}
%\tablewidth{0pt}
\tablehead{
\colhead{Element} & \colhead{Depletion\tablenotemark{a}}
}
\startdata
H  & 0.00 \\
He & 0.00 \\
C  & -0.30 \\
N  & -0.22 \\
O  & -0.22 \\
Ne & 0.00 \\
Na & -0.60 \\
Mg & -0.70 \\
Al & -1.70 \\
Si & -1.00 \\
S  & 0.00 \\
Cl & -0.30 \\
Ar & 0.00 \\
Ca & -2.52 \\
Fe & -2.00 \\
Ni & -1.40
\enddata
\tablenotetext{a}{Depletion given as
$\log(X/\mathrm{H})_{\mathrm{gas}}-
\log(X/\mathrm{H})_{\mathrm{ISM}}$}
\end{deluxetable}

\clearpage
%
%  V & O
%
\begin{figure}[!htp]
%\plottwo{f1a.eps}{f1b.eps}\\
\epsscale{2.25}
%\plottwo{f1c.eps}{f1d.eps}\\
\epsscale{4}
%\plottwo{f1e.eps}{f1f.eps}\\
\caption{\label{fig:VO} The standard optical line diagnostic
diagrams of \citet{VO87} (V\&O) showing [\nii]$\lambda 6583$/H$\alpha$,
[\oi]$\lambda 6300$/H$\alpha$ and [\sii]$\lambda\lambda
6717,30$/H$\alpha$ versus [\oiii]$\lambda 5007$/H$\beta$. The data
points are from (VO87) and curves are as labelled in the keys.}
\end{figure}

\newpage
%
% ADT
%
\begin{figure}[!htp]
\epsscale{1}
%\plottwo{f2a.eps}{f2b.eps}\\
\epsscale{2.25}
%\plottwo{f2c.eps}{f2d.eps}\\
\epsscale{4}
%\plottwo{f2e.eps}{f2f.eps}\\
\caption{\label{fig:ADT}UV diagnostics from \citet[][VTC97]{VTC97} and
\citet[][ADT]{ADT98}. Observations are from ADT, with asterisks
represent high-$z$ radio galaxies and triangles representing HST
observations of three Seyfert 2 NLRs. Dust-free curves are marked with
diamonds and dusty curves are marked with squares. Arrow represents
magnitude and direction of 3 visual 
magnitudes extinction of models through a dusty screen.}
\end{figure}

\clearpage
\begin{figure}[!htp]
\epsscale{1}
%\plottwo{f3a.eps}{f3b.eps}
\caption{\label{fig:ADTCC}UV diagnostic from ADT, demonstrating the
effects of metallicity on the dusty model (a) and differences between
the two models (b). $\alpha$ decreases from top to bottom for both
dusty (square) and dust-free (diamond) model curves in
(b). Observations (triangles) are from ADT. Reddening arrow as in
figure \ref{fig:ADT}a} 
\end{figure}

\begin{figure}[!htp]
%\plottwo{f4a.eps}{f4b.eps}
\caption{\label{fig:ADTNeO}UV diagnostic from ADT. Triangles represent
HST observations from ADT, asterisks represent observations of Seyfert
2s from \citet{Koski78} and crosses represent observations of Seyfert
2s from \citet{Allen98}. In (b) [\oiii]/H$\beta$ increases with
$\alpha$ for both model sets.}
\end{figure}

\clearpage
%
% Inskip
%
\begin{figure}[!htp]
%\plottwo{f5a.eps}{f5b.eps}
\caption{\label{fig:Inskip} UV diagnostic diagram from
\citet{Best00b,Inskip02a}. Observations from \citet{Inskip02a} with asterisks
representing 3C galaxies and triangles 6C galaxies. In (b) $\alpha$
decreases with increasing \ciii]/\cii]. Dusty model curves are marked with
squares and dust-free models with diamonds.} 
\end{figure}

%
% Other diagrams
\newpage
%
% UV
%
\begin{figure}[!htp]
%\plotone{f6.eps}
\caption{\label{fig:UV1}UV diagnostic diagram of a dusty model with
$Z= 2Z_\odot$ and $\alpha=-1.4$ as marked, showing curves of different
density as labelled. Lines of constant ionization parameter are as labelled.}
\end{figure}

\begin{figure}[!htp]
%\plotone{f7.eps}
\caption{\label{fig:UV2}UV diagnostic diagram. Labelled as in previous
diagram. This diagram demonstrates both a sensitivity to metallicity,
due to the component of secondary Nitrogen, and a sensitivity to
ionization parameter, revealing the stagnation due to radiation
pressure on dust.}
\end{figure}

\begin{figure}[!htp]
%\plotone{f8.eps}
\caption{\label{fig:UV3}UV diagnostic diagram. Labelled as in previous
diagram. Shows a strong metallicity sensitivity on both axis due to
Nitrogen in each ration.}
\end{figure}

\begin{figure}[!htp]
%\plotone{f9.eps}
\caption{\label{fig:UV4}UV diagnostic diagram. Labelled as in previous
diagram. Both metallicity and $U$ sensitive as in figure \ref{fig:UV2}.}
\end{figure}

%
% Optical
%
\clearpage
\begin{figure}[!htp]
%\plotone{f10.eps}
\caption{\label{fig:NeO}Near-UV optical diagnostic diagram showing
density variations on both the dusty (squares) and dust-free
(diamonds) models. Lines of constant are $U_0$ marked on both. Density
increases from top to bottom. Observations from \citet{Koski78}
(asterisks) and \citet{Allen98} (crosses). Reddening arrow indicates
magnitude and direction of 3 A$_V$ extinction on models.}
\end{figure}

\begin{figure}[!htp]
%\plotone{f11.eps}
\caption{\label{fig:OIIOIII}Optical diagnostic diagram suggested by
\citet{BPT81} showing $\alpha$ variation in dusty and dust-free
model. Curves and data marked as in previous diagram. Reddening arrow
indicates 10A$_V$ extinction here. $\alpha$ increases with [\oiii]/H$\beta$.}
\end{figure}

\begin{figure}[!htp]
%\plotone{f12.eps}
\caption{\label{fig:HeO}\citet{bws96} diagnostic diagram showing
$\alpha$ variation in dusty and dust-free models. Models and data
marked as in previous diagram. $\alpha$ increases from right to left.}
\end{figure}

\begin{figure}[!htp]
%\plotone{f13.eps}
\caption{\label{fig:Evansa} Diagnostic diagram showing metallicity
variations of dusty model. Curves and data marked as before.}
\end{figure}

\clearpage

\begin{figure}[!htp]
%\plotone{f14.eps}
\caption{\label{fig:ROIII} Temperature sensitive diagnostic
diagram showing density variation of both dusty and dust-free model
curves. Density increases left to right at $U_0=10^{-3}$. In addition
to the \citet{Koski78} (asterisks) and \citet{Allen98} (crosses) data,
data from \citet{Tadhunter89} is shown.}
\end{figure}

\begin{figure}[!htp]
%\plotone{f15.eps}
\caption{\label{fig:Evansb} Helium - Oxygen diagnostic diagram showing
density variations on dusty and dust free models. Density decreases
from top to bottom for both sets of model curves.}
\end{figure}

\begin{figure}[!htp]
%\plotone{f16.eps}
\caption{\label{fig:HeOIII}Helium - oxygen diagnostic diagram with
density variations of both models. Density increasing with increasing
[\oiii]/H$\beta$. Symbols as before.}
\end{figure}

\clearpage

\begin{figure}[!htp]
%\plottwo{f17a.eps}{f17b.eps}
\caption{\label{fig:HeNII}Helium - nitrogen diagnostic diagram showing
metallicity variation of dusty model (a) and density variations of
dusty and dust-free models (b). Symbols as before.}
\end{figure}

\newpage
\begin{figure}[!htp]
%\plotone{f18.eps}
\caption{\label{fig:NO}Density variations of both models in low
ionization species diagnostic diagram. Density increases right to
left for both sets of curves.}
\end{figure}

\begin{figure}[!htp]
%\plotone{f19.eps}
\caption{\label{fig:NIISII}Metallicity variation of dusty model on
nitrogen - sulfur diagnostic diagram.}
\end{figure}

\begin{figure}[!htp]
%\plotone{f20.eps}
\caption{\label{fig:OIISII} [\oii]$\lambda\lambda3727,9
\times$[\sii]$\lambda\lambda6717,30$
/[\sii]$\lambda\lambda4067,76\times$[\oii]$\lambda\lambda7318,24$
versus [\oiii]$\lambda 5007$/H$\beta$ diagnostic diagram showing
density variation of dusty model.}
\end{figure}

%
% IR
%
\clearpage
\begin{figure}[!htp]
%\plotone{f21.eps}
\caption{\label{fig:IR1} Infrared diagnostic diagram showing curves of
different $\alpha$ for the dusty model.}
\end{figure}

\begin{figure}[!htp]
%\plotone{f22.eps}
\caption{\label{fig:IR2}Comparison of the dusty (square) and dust-free
(diamond) models in the Infrared diagnostic diagram. Curves of
different $\alpha$ are shown, with $\alpha$ increasing from top to
bottom for both models. Note the different scales on each axis.}
\end{figure}

\end{document}